\documentclass[twocolumn,showpacs,aps,amsmath,amssymb,superscriptaddress,jcp]{revtex4-2}
\usepackage{adjustbox}
\usepackage{amsmath}
\usepackage{booktabs}
\usepackage{bm}
\usepackage{dcolumn}
\usepackage{hyperref}
\usepackage{graphicx}
\usepackage{siunitx}

\hypersetup{
    colorlinks=true,
    linkcolor=blue,
    urlcolor=black,
    citecolor=blue,
    pdfborder={0 0 0}  
}

\newcommand{\REF}{^{\rm ref}}

\newcommand{\tot}{_{\rm tot}}
\newcommand{\ext}{_{\rm ext}}
\newcommand{\ee}{_{\rm ee}}
\newcommand{\X}{_{\rm X}}
\newcommand{\C}{_{\rm C}}
\newcommand{\XC}{_{\rm XC}}

\newcommand{\BL}{^{\rm B3LYP}}
\newcommand{\ML}{^{\rm ML}}
\newcommand{\MLDFA}{^{\rm \text{ML\textendash DFA}}}
\newcommand{\MLBL}{^{\rm \text{ML\textendash B3LYP\textendash p}}}

\begin{document}


\title{Mitigating error cancellation in density functional approximations via machine learning correction}


\author{Zipeng~An}
\affiliation{Hefei National Research Center for Physical Sciences at the Microscale
\& Synergetic Innovation Center of Quantum Information and Quantum Physics,
University of Science and Technology of China, Hefei, Anhui 230026, China}

\author{JingChun~Wang} 
\affiliation{Department of Chemistry, University of Basel, Klingelbergstrasse 80, CH-4056 Basel, Switzerland}

\author{Yapeng~Zhang}
\affiliation{Hefei National Research Center for Physical Sciences at the Microscale
\& Synergetic Innovation Center of Quantum Information and Quantum Physics,
University of Science and Technology of China, Hefei, Anhui 230026, China}

\author{Zhiyu~Li}
\affiliation{Department of Chemistry, Fudan University, Shanghai 200438, China}

\author{Jiang~Wu}
\affiliation{Department of Chemistry, The University of Hong Kong, Pokfulam Road, Hong Kong, China}

\author{Yalun~Zheng}
\affiliation{Department of Chemistry, The University of Hong Kong, Pokfulam Road, Hong Kong, China}

\author{GuanHua~Chen} \email{ghc@everest.hku.hk}
\affiliation{Department of Chemistry, The University of Hong Kong, Pokfulam Road, Hong Kong, China}

\author{Xiao~Zheng} \email{xzheng@fudan.edu.cn}
\affiliation{Department of Chemistry, Fudan University, Shanghai 200438, China}


\date{Submitted on February~26, 2025}

\begin{abstract}

The integration of machine learning (ML) with density functional theory has emerged as a promising strategy to enhance the accuracy of density functional methods. While practical implementations of density functional approximations (DFAs) often exploit error cancellation between chemical species to achieve high accuracy in thermochemical and kinetic energy predictions, this approach is inherently system-dependent, which severely limits the transferability of DFAs. To address this challenge, we develop a novel ML-based correction to the widely used B3LYP functional, directly targeting its deviations from the exact exchange-correlation functional. By utilizing highly accurate absolute energies as exclusive reference data, our approach eliminates the reliance on error cancellation. To optimize the ML model, we attribute errors to real-space pointwise contributions and design a double-cycle protocol that incorporates self-consistent-field calculations into the training workflow. Numerical tests demonstrate that the ML model, trained solely on absolute energies, improves the accuracy of calculated relative energies, demonstrating that robust DFAs can be constructed without resorting to error cancellation. Comprehensive benchmarks further show that our ML-corrected B3LYP functional significantly outperforms the original B3LYP across diverse thermochemical and kinetic energy calculations, offering a versatile and superior alternative for practical applications.

\end{abstract}

\maketitle


\section{Introduction} \label{sec:intro}

Density functional theory (DFT) has become a widely used tool for studying the electronic structure of matter \cite{hohenberg1964inhomo, kohn1965self, parr1995density, engel2011density, jones2015density, narbe2017thirty}. Despite its significant advantages, DFT methods have generally not achieved chemical accuracy, particularly for certain complex systems, due to the unknown form of the exact exchange-correlation (XC) functional. To tackle this challenge,  various density functional approximations (DFAs) have been proposed, introducing more sophisticated electron density descriptors that enhance computational accuracy, albeit at the expense of efficiency. This progression is often illustrated through the concept of ``Jacob's Ladder'' \cite{perdew2001jacob, perdew2005prescription}. Among the numerous DFAs, the B3LYP functional, which adopts the hybrid form of Becke \cite{becke1988density, becke1993density} with the correlation functional of Lee, Yang and Parr \cite{lee1988development}, remains one of the most popular due to its excellent balance between accuracy and efficiency \cite{Stephens1994AbIC}. Thus, a core challenge for DFT today is to develop a DFA that performs uniformly better than B3LYP in terms of both efficiency and accuracy \cite{Cohen2012challenges}.

With the rapid advancement of artificial intelligence, machine learning (ML) techniques \cite{bishop2006pattern, marsland2014machine, jordan2015machine} such as neural networks (NNs) and decision trees have been employed to improve the performance of DFT methods. For instance, ML models have been utilized to explicitly establish complex mappings, such as the Hohenberg-Kohn mapping \cite{hohenberg1964inhomo} and the Kohn-Sham mapping \cite{kohn1965self}, within the DFT framework \cite{tozer1996exchange, zheng2004generalized, wang2004improving, li2007improving, snyder2012finding, bart2013machine, snyder2013orbital, sun2014alternative, li2016understanding, li2016pure, brockherde2017bypassing, liu2017improving, nagai2018neural, seino2018semilocal, zhou2019toward, lin2019classical, schmidt2019machine, ryczko2019deep, lei2019design, nagai2020completing, fujinami2020orbital, vargas2020bayesian, moreno2020deep, denner2020efficient, meyer2020machine, kirkpatrick2021pushing, margraf2021pure, bhattacharjee2021regularized, gedeon2021machine, alghadeer2021highly, cuierrier2021constructing, li2021kohn,zheng2021artificial, yang2022predicting, nagai2022machine, pokharel2022exact, liu2022supervised, wu2023construct, wu2023redesigning, liu2023supervised, chen2024development, wang2024toward}. Moreover, ML algorithms have been applied to enhance the accuracy of self-consistent-field (SCF) density functional calculations \cite{dick2020machine, chen2021deepks}, and to construct post-SCF corrections \cite{hu2003combined, wang2004combined, duan2004accurate, li2007improving, wu2007x1, wu2008improving, wu2009accurate, wu2010x1s, zhou2016improving}. The latter approach can be categorized into $\Delta$-ML methods \cite{hu2003combined, sun2014alternative, ramakrishnan2015big}.

Recently, Wang et~al. have utilized ML models to construct corrections for the XC energy given by several existing DFAs \cite{wang2022improving, wang2023semilocal}. The ML-based energy corrections typically exhibit a small magnitude, allowing the corrected DFA to largely preserve the physical constraints or conditions imposed by the original DFA. This enables the ML-corrected DFA to outperform its parent functional without compromising its inherent advantages. Specifically, ML models were employed to alleviate errors in the thermochemical and kinetic energies calculated using the Perdew-Burke-Ernzerhof (PBE) functional \cite{perdew1996generalized} and the hybrid B3LYP functional, diminishing their discrepancies from highly accurate reference data.
The resulting ML-PBE \cite{wang2022improving} and ML-B3LYP functionals \cite{wang2023semilocal} demonstrate significantly improved accuracy in predicting atomization energies. Additionally, ML-B3LYP allows for conducting SCF calculations with similar efficiency to the original B3LYP functional.

Despite the promising performance of the ML-corrected functionals developed in Refs.~\cite{wang2022improving, wang2023semilocal}, they show little to marginal improvement over the original DFAs beyond atomization energies. For example, ML-B3LYP maintains the same level of accuracy for isomerization energies and reaction barrier heights as the original B3LYP functional \cite{wang2023semilocal}.
The unsatisfactory generalization ability of ML corrections is likely due to two aspects of the training procedure. First, the training data and targets primarily consist of energy differences between chemical species. Consequently, placing emphasis on relative energies during the training process makes the resulting ML model highly dependent on the cancellation of errors among chemical species. Second, the reference data used for training are limited, covering only a small fraction of the chemical space, which restricts the transferability of the ML model. 

The goal of this work is to address these challenges and enhance the generalization capability of ML corrections. We focus on the widely adopted B3LYP hybrid functional, formally expressed as:
\begin{align}
    E\XC\BL [\rho] &= a_0 E\X^{\rm LSDA} [\rho] + (1 - a_0) E\X^{\rm HF} [\rho] + a\X \Delta E\X^{\rm B88} [\rho] \\ \nonumber
    & \quad + a\C E\C^{\rm LYP} [\rho] + (1 - a\C) E\C^{\rm VWN} [\rho],
\end{align}
where $E\X^{\rm LSDA}$ and $E\X^{\rm HF}$ denote the local spin density \cite{becke1989density} and Hartree-Fock exchange functionals, respectively, $\Delta E\X^{\rm B88}$ represents the difference between the Becke88 \cite{becke1988density} and local spin density exchange functionals, and $E\C^{\rm LYP}$ and $E\C^{\rm VWN}$ correspond to the Lee-Yang-Parr \cite{lee1988development} and Vosko-Wilk-Nusair correlation functionals \cite{vosko1980accurate}. The hybrid coefficients $a_0$, $a\X$, and $a\C$, originally optimized via semi-empirical fitting to experimental or high-accuracy computational data \cite{becke1993density}, govern the relative contributions of these components. While these coefficients are fixed in the standard B3LYP functional \cite{Stephens1994AbIC}, they are inherently electron density-dependent and thus system-specific \cite{zheng2004generalized, wu2023construct}. In principle, the exact XC functional can be formally realized by reformulating these coefficients as functionals of electron density, i.e., $a_0 [\rho]$, $a\X [\rho]$, and $a\C [\rho]$. Consequently, the deviation of B3LYP from the exact functional can be characterized by a density-dependent correction term \cite{wu2023construct}.

To address this deviation, we employ an ML model to represent the correction term bridging B3LYP and the exact XC functional. While traditional DFAs rely on error cancellation between chemical species to achieve high accuracy in thermochemical and kinetic energy predictions, such system-dependent strategies inherently limit the transferability of DFAs. By contrast, our ML-based correction eliminates the need for error cancellation by focusing exclusively on absolute energies during the training process. Specifically,  highly accurate training data are generated using coupled-cluster theory with single, double, and perturbative triple excitations (CCSD(T)) \cite{Krishnan1989fifthorder, Bartlett2007cc}, which is widely regarded as the gold standard in quantum chemistry. 



Additionally, to develop an ML-based mapping from electron density to XC energy density that is applicable across a wide range of chemical species, we choose to sample within the space of electron density rather than directly exploring the vast chemical space. To achieve this, we must provide accurate reference values for numerous grid points in real space, resulting in a pointwise ML correction model, which we will refer to as ML-B3LYP-p. In contrast, the ML-corrected B3LYP functional developed in Ref.~\cite{wang2023semilocal} will be designated as ML-B3LYP-g, since that model was intended to correct the global energies of chemical species. 

In this work, we provide extensive numerical examples which demonstrate that the ML-B3LYP-p functional consistently outperforms the conventional B3LYP functional across a wide range of energetic properties, establishing it as an immediate and superior alternative.

The remainder of this paper is organized as follows: Section~\ref{sec2} introduces the design and training strategy of the pointwise ML model for correcting the B3LYP functional. Section~\ref{discussion} presents the performance of the resulting ML-B3LYP-p functional across various datasets. Finally, Section~\ref{sec:conclusion} gives the conclusion and outlook.

\section{Methodology} \label{sec2}

\subsection{Design of ML-corrected DFA} \label{subsec:design}

Instead of constructing an ML-based DFA from the ground up, we develop an ML model to correct an existing DFA. The XC energy of the ML-corrected functional is given by  
%
\begin{equation}
 E_{\text{XC}}\MLDFA = E_{\text{XC}}^{\mathrm{DFA}} + \int \mathrm{d}\mathbf{r}\, \rho (\mathbf{r}) \Delta \epsilon_{\text{XC}}^{\text{ML}} (\mathbf{r}).
    \label{ML-DFA}
\end{equation}
Here, $\rho (\mathbf{r})$ denotes the electron density, and $\Delta \epsilon_{\text{XC}}^{\text{ML}} (\mathbf{r})$ represents the ML correction to the XC energy density per electron. 
Throughout this work, we choose the B3LYP functional that incorporates the VWN5 correlation \cite{vosko1980accurate} as the parent DFA.

Following Ref.~\cite{wang2023semilocal}, we establish a semilocal mapping that associates $\Delta \epsilon_{\text{XC}}^{\text{ML}}$ at a point $\mathbf{r}$ with the density descriptors evaluated at that point. Here, ``semilocal'' indicates that a density descriptor may depend on the gradient of the electron density. 
The density descriptors employed include the Wigner-Seitz radius $r_{s} = (4\pi\rho/3)^{-1/3}$, the relative spin polarization $\zeta = (\rho_{\downarrow} - \rho_{\uparrow})/\rho$, and the reduced density gradient $s = \frac{|\nabla \rho|}{2(3 \pi ^{2})^{1/3} \rho^{4/3}}$ \cite{perdew1996generalized}. 
In addition to these generalized gradient approximation (GGA) level descriptors, we also incorporate three descriptors related to kinetic energy density: 
$z = \tau^{\rm W}/\tau$, $\alpha = (\tau - \tau^{\rm W})/\tau^{\rm unif}$, and $t^{-1} = \tau/\tau^{\rm unif}$, thus extending the semilocal mapping to the meta-GGA level \cite{sun2013density}. 
Here, $\tau = \frac{1}{2} \frac{\sum_{i=1}^{\mathrm{occ}} \left| \nabla \phi_{i} \right|^2}{\rho(\mathbf{r})}$,
where $\phi_{i}$ represents the $i$th Kohn-Sham orbital. The term $\tau^{\rm W} = \frac{1}{8} \frac{|\nabla \rho|^2}{\rho}$ denotes the von~Weizs\"{a}cker kinetic energy density, and $\tau^{\rm unif} = (3/10)(3 \pi ^{2})^{2/3} \rho^{5/3}$ represents the kinetic energy density of a uniform electron gas \cite{parr1995density}. Altogether, the semilocal mapping can be expressed as
\begin{equation}
\{r_{s}, \zeta, s, z, \alpha, t^{-1}\} \mapsto \Delta \epsilon_{\text{XC}}^{\text{ML}}(\mathbf{r}).
\label{map}
\end{equation}
%
%
The XC potential for a spin-$\sigma$ electron is evaluated as
\begin{equation}
v_{{\rm XC}; \sigma} \MLDFA \equiv 
\frac{\delta E_{\rm XC}\MLDFA[\rho]}{\delta \rho_{\sigma}} = 
v_{{\rm XC}; \sigma}^{\rm DFA} + \Delta v_{{\rm XC}; \sigma}^{\rm ML}
\label{xc-potential}
\end{equation}
where $\sigma = \uparrow, \downarrow$, and
\begin{equation} \label{dvxc-1}
\Delta v^{\mathrm{ML}}_{\mathrm{XC};\sigma}(\mathbf{r}) = \frac{\partial f^{\mathrm{ML}}}{\partial \rho_{\sigma}} - \nabla\cdot\frac{\partial f^{\mathrm{ML}}}{\partial\nabla\rho_{\sigma}} + \frac{\partial f^{\mathrm{ML}}}{\partial \tau_{\sigma}} \frac{\partial \tau_{\sigma}}{\partial \rho_{\sigma}}
\end{equation}
with $f^{\text{ML}} \equiv \rho\,\Delta\epsilon^{\text{ML}}_{\rm XC}$.

\begin{figure}[t]
  \centering
  \includegraphics[width=\columnwidth]{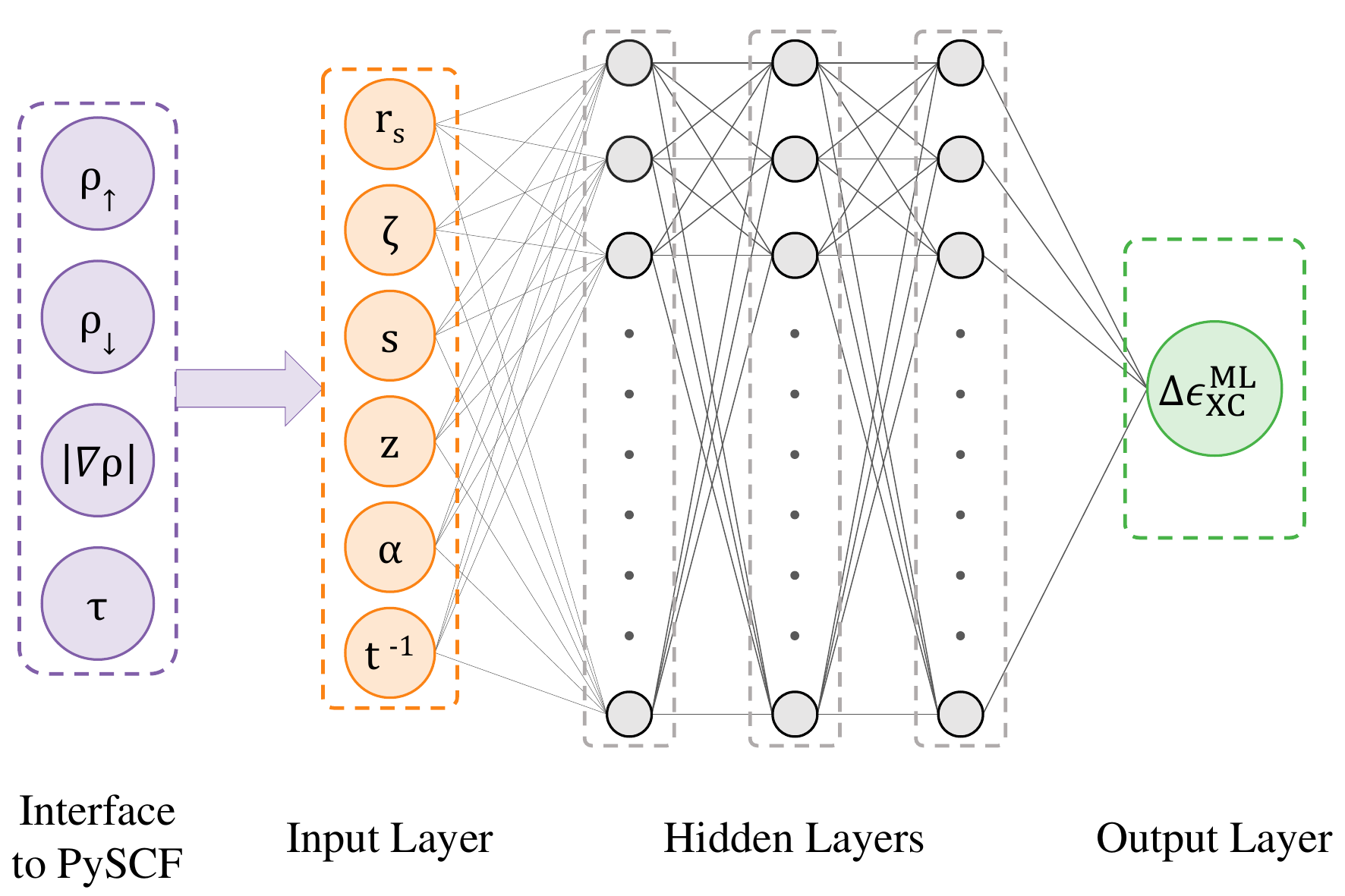} 
  \caption{Schematic illustration of the NN used for correcting the XC energy density. The basic density variables, $\{\rho_{\uparrow}, \rho_{\downarrow}, \lvert \nabla \rho \rvert, \tau\}$, are evaluated using the quantum chemistry software PySCF \cite{sun2015libcint, sun2018pyscf, sun2020recent}. These variables are then transformed into the semilocal density descriptors $\{r_{s}, \zeta, s, z, \alpha, t^{-1}\}$ and input into a fully connected NN. The NN consists of three hidden layers, each containing 40 hidden neurons. The output layer comprises a single neuron that produces $\Delta\epsilon^{\text{ML}}_{\rm XC}$. }
  \label{fig1}
\end{figure}

As illustrated in Fig.~\ref{fig1}, the construction of the NN begins by obtaining the basic density variables $\{\rho_{\uparrow}, \rho_{\downarrow}, \lvert \nabla \rho \rvert, \tau\}$ for each pointwise sample. These samples correspond to the $\mathbf{r}$ points in real space, generated using PySCF, an open-source quantum chemistry software \cite{sun2015libcint, sun2018pyscf, sun2020recent}, for density functional calculations. For each $\mathbf{r}$ point, the basic density variables are transformed into six semilocal descriptors, $\{r_{s}, \zeta, s, z, \alpha, t^{-1}\}$, which are then fed into the input layer of the NN. 
The architecture of the NN is configured as $6 \times 40 \times 40 \times 40 \times 1$, comprising six input neurons, three hidden layers (each containing 40 neurons), and one output neuron that produces the final energy density correction.  All components of the NN are implemented using the PyTorch platform \cite{paszke2019pytorch}.
Additionally, the partial derivatives $\frac{\partial f^{\mathrm{ML}}}{\partial \rho_{\sigma}}$, $\frac{\partial f^{\mathrm{ML}}}{\partial\nabla\rho_{\sigma}}$, and $\frac{\partial f^{\mathrm{ML}}}{\partial \tau_{\sigma}}$, required for computing $\Delta v^{\mathrm{ML}}_{\mathrm{XC};\sigma}(\mathbf{r})$ via Eq.~\eqref{dvxc-1}, are calculated using the automatic differentiation protocol of PyTorch.

\subsection{Training strategy and reference data} \label{subsec:training}

\begin{figure*}[t]
  \centering
  \includegraphics[width=\textwidth]{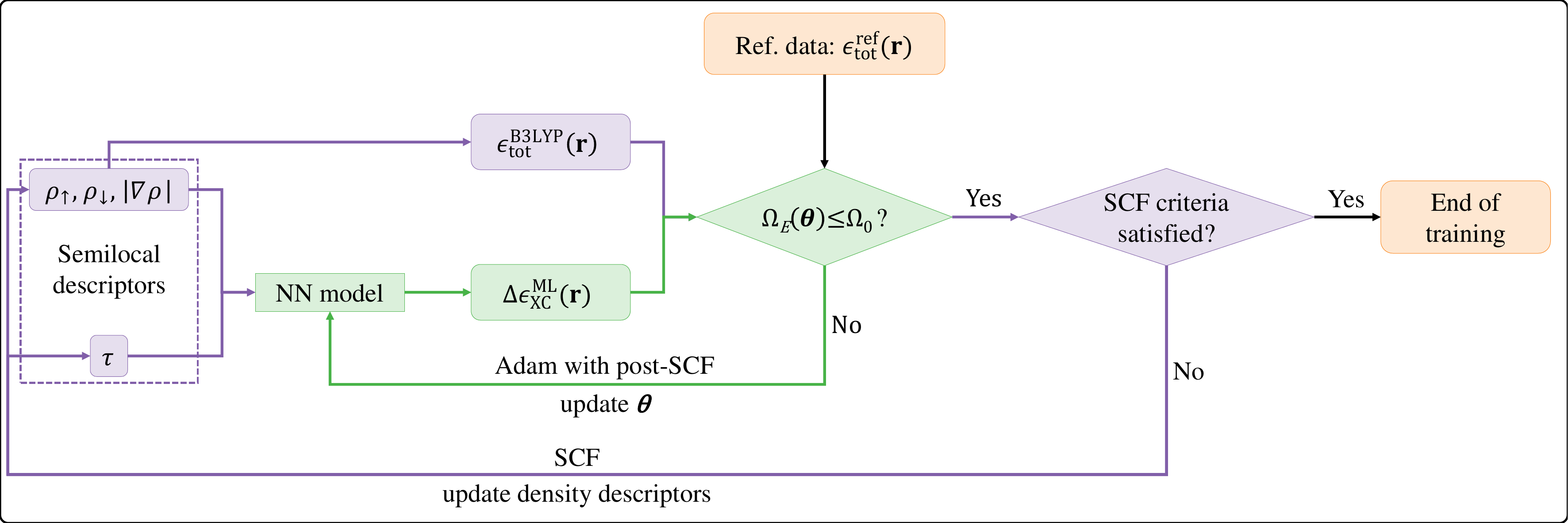} 
  \caption{Schematic diagram illustrating a double-cycle workflow for training the ML-B3LYP-p functional. Specifically, the inner and outer cycles are marked in green and purple, respectively. Related details are provided in Section~\ref{subsec:training}. The parameter $\Omega_0$ denotes the predefined threshold for minimizing the loss function $\Omega_E$.}
  \label{fig2}
\end{figure*}

The training procedure for the NN model follows the strategy outlined in Ref.~\cite{wang2023semilocal}, while incorporating several important modifications to enhance the generalization ability of the ML-B3LYP-p functional.

As mentioned in Section~\ref{sec:intro}, previously developed ML-corrected DFAs primarily utilized relative energies to train the ML models \cite{wang2022improving, wang2023semilocal}. This approach inevitably leads to unsatisfactory generalization ability of the ML models due to its reliance on error cancellation. In contrast, our new training strategy exclusively employs absolute energies of atoms and molecules. Consequently, the resulting ML-corrected functional is expected to avoid error cancellation among chemical species and demonstrate substantially enhanced generalization capability by predicting accurate absolute energies, as well as various types of relative energies.

A pointwise training protocol was proposed in Ref.~\cite{wang2022improving}, where reference values for numerous $\mathbf{r}$ points are obtained by decomposing the global energy error of a chemical species into contributions from each $\mathbf{r}$ point. This approach assumes that the magnitude of the pointwise error is proportional to the magnitude of the XC energy density itself. However, because these pointwise errors originate from the same global error through a simple rule, they are intrinsically correlated and do not adequately represent the desired mutually independent samples. 

In our new strategy, the energy densities generated by SCF calculations using the CCSD(T) method serve as the reference data. For a given chemical species, the pointwise energy density at any $\mathbf{r}$ point comes from the same second-order reduced density matrix, which typically has a sufficiently large dimension to ensure that these pointwise data are effectively independent. Below, we provide the detailed formulas for the evaluation of pointwise reference data.

For a given chemical species, the reference value for its total (or absolute) energy, $E\tot\REF$, is obtained through SCF calculation using a highly accurate quantum chemistry method. This value can be expressed as an integral over the energy density per electron, $\epsilon\tot\REF(\mathbf{r})$, as follows,
\begin{equation} \label{energy_ee}   
E\tot\REF = \int  \rho\REF(\mathbf{r})\,  \epsilon\tot\REF(\mathbf{r}) \, \mathrm{d} \mathbf{r}.
\end{equation}
%
%
The definition of $\epsilon\tot\REF(\mathbf{r})$ is inherently non-unique due to the flexibility in choosing the ``gauge''. Specifically, any function $\Delta \epsilon(\mathbf{r})$ can be added to $\epsilon\tot\REF(\mathbf{r})$ as long as the condition $\int  \rho\, \Delta \epsilon \, \mathrm{d} \mathbf{r} = 0$ is satisfied. To proceed, we select a simple gauge that allows the energy density to be further decomposed into several components:
\begin{equation} \label{energy_density_tot}
\epsilon_{\rm tot}\REF (\mathbf{r}) = \tau\REF (\mathbf{r}) + \epsilon\ext(\mathbf{r}) + \epsilon\ee\REF (\mathbf{r}),
\end{equation}
where
\begin{align}
\tau\REF(\mathbf{r}) &= \frac{1}{2} \frac{\sum_{i=1}^{\mathrm{occ}} \left| \nabla \psi_{i} \right|^2}{\rho\REF(\mathbf{r})},   \\
\epsilon\ext(\mathbf{r}) &= -\sum_{A} \frac{Z_A}{\left\vert \mathbf{r} - \mathbf{R}_A \right\vert},\\
\epsilon\ee\REF (\mathbf{r}) &= \frac{1}{2}\sum_{i,j}\frac{\psi_i^*(\mathbf{r})\psi_j^*(\mathbf{r})}{\rho\REF(\mathbf{r})} \sum_{k,l} P\REF_{ij,kl}  \nonumber \\
 & \quad \times \int\frac{\psi_k(\mathbf{r'})\psi_l(\mathbf{r'})}{\left| \mathbf{r} - \mathbf{r'} \right|}\mathrm{d}\mathbf{r'}.
\end{align}
Here, $\psi_i$ represents the $i$th Hartree-Fock orbital, $P\REF_{ij,kl}$ denotes an element of the second-order reduced density matrix, and $Z_A$ and $\mathbf{R}_A$ represent the charge and position of the nucleus $A$, respectively.

In contrast, the energy density of the ML-B3LYP-p functional can be expressed as
\begin{equation} \label{te_ml}
E\tot\MLBL = \int \rho(\mathbf{r})\, \epsilon\tot\MLBL(\mathbf{r}) \, \mathrm{d} \mathbf{r},
\end{equation}
where
\begin{align}
\epsilon_{{\rm tot}}\MLBL (\mathbf{r}) &= \tau(\mathbf{r}) + \epsilon_{{\rm ext}}(\mathbf{r}) + \epsilon_{J}(\mathbf{r}) + \epsilon_{{\rm XC}}\BL (\mathbf{r}) \nonumber \\
&\quad  + \Delta \epsilon_{{\rm XC}}\ML (\mathbf{r}).
\label{te_den_ml}
\end{align}
Here, $\epsilon_{J}(\mathbf{r})$ represents the classical electron-electron Coulomb repulsion energy, while $\epsilon_{{\rm XC}}\BL (\mathbf{r})$ corresponds to the XC energy of the conventional B3LYP functional.

The training process aims to minimize the energy loss function, defined as
\begin{align}
\Omega_{E}(\bm{\theta}) &= \sum_{a} w_{a}^{\rm data} \Big\lvert 
\sum_{i} w_{ai}^{\rm grid} \Big[ \epsilon_{{\rm tot}}\REF (\mathbf{r}_{ai}) \, \rho\REF (\mathbf{r}_{ai}) \nonumber \\
&\quad - \epsilon_{{\rm tot}}\MLBL (\mathbf{r}_{ai}) \, \rho\MLBL (\mathbf{r}_{ai}) \Big] \Big\rvert.
\label{lossfunction}
\end{align}
Here, $a$ labels the chemical species in the training and validation datasets, $\mathbf{r}_{ai}$ represents the $i$th grid point of the $a$th species, $\bm\theta$ denotes the set of NN parameters, and ${w}_{a}^{\rm data}$ and ${w}_{ai}^{\rm grid}$ represent the weights associated with the occurrence frequency of a chemical species in the dataset and the relative significance of a pointwise data sample, respectively.
Specifically, the weights ${w}_{a}^{\rm data}$ are introduced to account for the relative significance of different chemical species, allowing us to place greater emphasis on common atoms such as hydrogen and carbon. Since these atoms often appear more frequently in the dataset, assigning larger weights to them effectively reduce the error accumulation. In practice, the values of ${w}_{a}^{\rm data}$ are determined based on their frequency of occurrence in the G2/97 dataset \cite{curtiss1997assessment, curtiss1998assessment}. 

Another limitation of the training strategy proposed in Ref.~\cite{wang2023semilocal} is the reliance on a non-gradient-based algorithm for optimizing the NN parameters $\bm\theta$. 
This limitation arises because the SCF process functions as a black box within the training workflow, preventing the backpropagation of errors through the NN. Consequently, automatic differentiation algorithms available in modern ML tools, as well as gradient-based optimizers, cannot be applied.

To enhance training performance by utilizing the gradients of pointwise errors, we design a double-cycle workflow for the training process, as illustrated in Fig.~\ref{fig2}. Specifically, in the inner cycle (marked in green), both $\Delta\epsilon^{\text{ML}}_{\rm XC}$ and $\Omega_E(\bm\theta)$ are calculated in a post-SCF manner. Consequently, gradient-based algorithms, such as adaptive moment estimation (Adam) \cite{Kingma2014vow}, can be employed to optimize the NN parameters. Meanwhile, in the outer cycle (marked in purple), the density descriptors are updated until the SCF criteria are satisfied. 

During the training process, the outer cycle is accessed after every 2000 iterations of the inner cycle. As a result, the SCF calculations account for only a small fraction of the total computational time, significantly enhancing the numerical efficiency of the training process.

\subsection{Datasets} \label{subset:dataset}

We select 11 small molecules, $\{\rm LiH$, $\rm CH$, $\rm NH$, $\rm NH_2$, $\rm NH_3$, $\rm OH$, $\rm HF$, $\rm SiH_2 (singlet)$, $\rm SiH_2 (triplet)$, $\rm SiH_3$, $\rm O_2\}$, from the G2/97 dataset, along with 18 atoms ranging from hydrogen (H) to argon (Ar), to form the training set. Moreover, we select another 10 molecules, $\{\rm BeH$, $\rm CH_2 (singlet)$, $\rm CH_2 (triplet)$, $\rm CH_3$, $\rm CH_4$, $\rm H_2O$, $\rm SiH_4$, $\rm PH_2$, $\rm PH_3$, $\rm H_2S\}$, to create the validation set, which helps prevent overfitting during training.

All absolute energies and pointwise energy densities are calculated using the CCSD(T) method with the def2-QZVPD \cite{rappoport2010a, weigend2003a} basis set. For density functional calculations, we employ the def2-TZVPD \cite{weigend2005a, rappoport2010a} basis. The number of pointwise samples ($\mathbf{r}$ points) is approximately 2.17 million for the training set and 2.54 million for the validation set. Due to limited computational resources, the molecules included in both sets primarily consist of hydrides, as the computation of the second-order reduced density matrix $P^{\rm ref}_{ij,kl}$ at the CCSD(T) level requires substantial computer memory. To enhance diversity, both the training and validation sets include closed-shell and open-shell molecules, as well as some free radicals.

We select two databases as test sets to evaluate the prediction capabilities of the ML-B3LYP-p functional. The first database, designated as the TEST12 database, comprises 12 datasets used in Ref.~\cite{wang2023semilocal} to assess the performance of the ML-B3LYP-g functional. The TEST12 database encompasses a diverse range of relative energies among chemical species, including atomization energies, enthalpies of formation, ionization potentials, electron affinities, isomerization energies, reaction barrier heights, and atom transfer energies. This diversity enables a direct and unbiased comparison between ML-B3LYP-p and the previously developed ML-B3LYP-g.

Specifically, the 12 datasets are: G2-AE (atomization energies of atoms and small molecules in the G2/97 dataset, which extends the original G2 test set and consists of 148 molecular systems) \cite{curtiss1997assessment, curtiss1998assessment}, G3-HOF (enthalpies of molecules in the G3/99 dataset, a benchmark set based on the G3 theoretical method, comprising 223 molecules) \cite{curtiss2005g3}, P6-AE (atomization energies of platonic hydrocarbons cages containing up to 20 carbon atoms) \cite{narbe2017thirty}, AlkAtom19-AE (atomization energies of linear alkanes containing up to 8 carbon atoms) \cite{narbe2017thirty}, G2-IP (adiabatic ionization potentials of atoms and small molecules in the G2/97 dataset) \cite{curtiss1997assessment, curtiss1998assessment}, G2-EA (adiabatic electron affinities of atoms and small molecules in the G2/97 dataset) \cite{curtiss1997assessment, curtiss1998assessment}, ISO-C (isomerization energies of the ground-state configurations of $\mathrm{C}_{20}$ and $\mathrm{C}_{24}$) \cite{narbe2017thirty}, ISO20 (isomerization energies involving molecules with $1-4$ non-hydrogen atoms) \cite{narbe2017thirty}, BDE99 (99 bond dissociation energies small molecules) \cite{narbe2017thirty}, BDE42 (42 bond dissociation energies of hydrocarbon and hydrocarbon derivative systems) \cite{wang2023semilocal}, HTBH38 (38 hydrogen transfer barrier heights) \cite{zhao2006new, zhao2005benchmark}, and NHTBH38 (38 non-hydrogen transfer barrier heights) \cite{zhao2006new, zhao2005benchmark}.

The second test set is the GMTKN55 database \cite{goerigk2017look}, which has been widely employed for benchmarking the performance of DFAs. The GMTKN55 database consists of 55 subsets, comprising 1505 relative energies and 2462 single-point energies. These relative energies are categorized into five groups: Category 1 includes basic properties and reaction energies for small systems; Category 2 covers reaction energies for large systems and isomerization reactions; Category 3 pertains to reaction barrier heights; Category 4 focuses on intermolecular noncovalent interactions (NCIs); and Category 5 addresses intramolecular NCIs.

The TEST12 and GMTKN55 databases contain some overlapping chemical species. However, this overlap does not hinder the evaluation of the ML model, as long as the two databases are assessed independently.

In Ref.~\cite{goerigk2017look}, two types of statistical errors, WTMAD-1 and WTMAD-2, were proposed to quantitatively assess the accuracy of a DFA across a comprehensive database covering various categories of energy data. Specifically, we choose to apply WTMAD-2 to evaluate the performance of the ML-corrected functionals. The WTMAD-2 of a database consisting of $N$ subsets is defined as follows,
\begin{equation}
    \text{WTMAD-2} = \frac{1}{\sum^{N}_{j=1} N_j} \sum^{N}_{j=1} N_j \cdot \frac{\overline{\left| \Delta E \right|}}{\overline{\left| \Delta E \right|}_j} \cdot \text{MAD}_j,  \label{wtmad2}
\end{equation}
where $j$ labels a subset within the database, $N_j$ denotes the number of single-point energies within the $j$th subset, and $\overline{|\Delta E|}_j$ and $\text{MAD}_j$ represent the averaged magnitude and mean absolute deviation (MAD) of these energies, respectively. Meanwhile, $|\overline{\Delta E}|$ refers to the average magnitude of all single-point energies in the entire database.

\section{Results and discussion} \label{discussion}

\subsection{Training performance} \label{subsec:train}

\begin{figure}
  \centering
  \includegraphics[width=\linewidth]{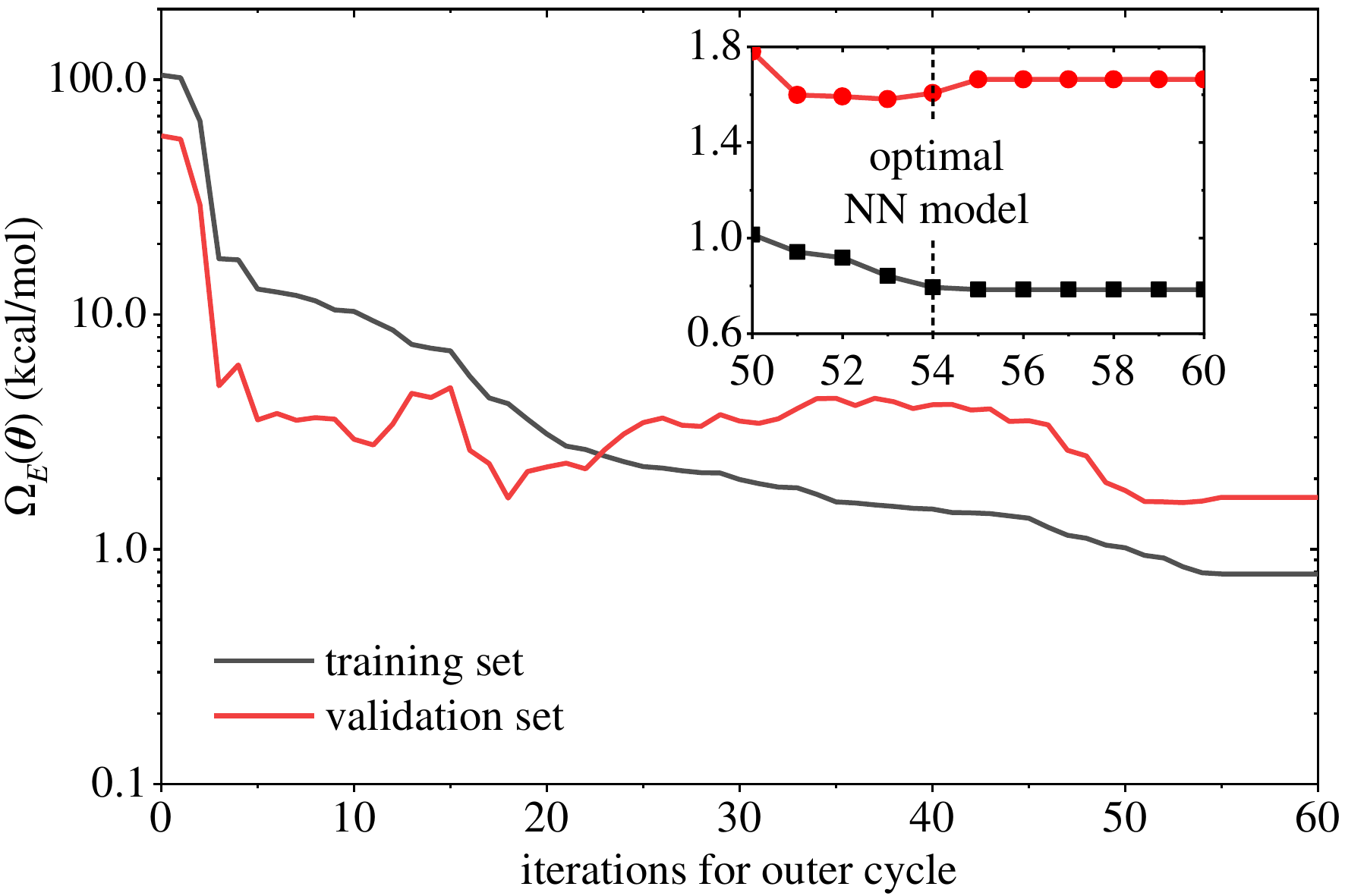} 
  \caption{Evolution of $\Omega_E(\bm\theta)$ during the training process. The inset highlights the region where the optimal NN model is selected for further testing. }
  \label{fig3}
\end{figure}

The training process is illustrated in Fig.~\ref{fig3}. The optimal NN model is selected based on the lowest loss for the validation set. Specifically, $\Omega_E$ is significantly reduced from 104.86 kcal/mol to 0.84 kcal/mol for the training set, and from 57.63 kcal/mol to 1.58 kcal/mol for the validation set. These systematically reduced deviations provide a solid foundation for enhancing the accuracy of the resulting ML-B3LYP-p functional.

With the ML correction, the MAD of total atomic energies for H--Ar reduces to 0.89 kcal/mol, a substantial decrease from the original MAD of 66.88 kcal/mol using the uncorrected B3LYP functional. Specifically, the deviation of hydrogen's total atomic energy from the reference value decreases from 0.77 kcal/mol to less than 0.01 kcal/mol, while the deviation for carbon decreases from 18.92 kcal/mol to 0.05 kcal/mol. The deviations for N, O, F, and Cl are also significantly reduced. 

The superior performance of the ML correction in predicting absolute energies is further demonstrated by calculating the total energies of the chemical species in the G2/97 dataset. The ML-B3LYP-p functional achieves an MAD of 2.5 kcal/mol compared to the reference values, drastically lower than the MAD of 111.89 kcal/mol obtained with the uncorrected B3LYP functional. In comparison, the ML-B3LYP-g functional developed in Ref.~\cite{wang2023semilocal}  yields an MAD of 8.3 kcal/mol. Thus, the remarkably low MAD achieved by the ML-B3LYP-p functional indicates its promising potential to eliminate the systematic errors in absolute energies, which are crucial for accurately predicting relative energies.


To ensure that the enhanced accuracy in energy prediction achieved by the ML correction does not compromise the accuracy of electron density \cite{medvedev2017density}, we also examine the electron density produced by SCF calculations using the ML-B3LYP-p functional, even though the density itself is not explicitly included in the training target. The error in electron density is quantified using the density loss function, defined as
\begin{equation} \label{delta_density}
\Omega_{\rho}(\boldsymbol{\theta}) = \sum_{a} \sum_i w_{ai}^{\rm grid}  \Big\vert \rho\REF(\mathbf{r}_{ai}) - \rho\MLBL (\mathbf{r}_{ai}) \Big\rvert. 
\end{equation}
We find that $\Omega_\rho$ remains nearly unchanged when applying the optimal NN model to both the training and validation sets. This confirms that the ML correction for energy does not sacrifice the quality of electron density.

\begin{figure*}[t]
  \centering
  \includegraphics[width=\textwidth]{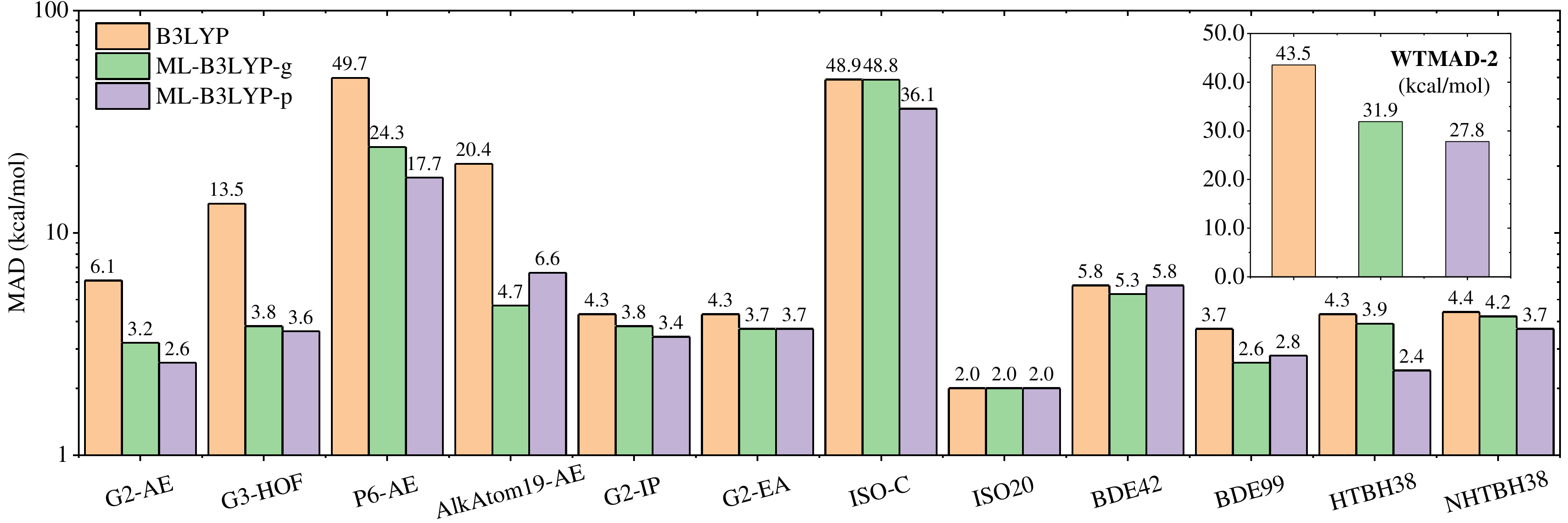} 
  \caption{Performance of the ML-B3LYP-p functional on the TEST12 database compared to the B3LYP and ML-B3LYP-g methods. The number above each bar indicates the MAD for the corresponding dataset resulting from the specific DFA; refer to the main text for further details. The optimized geometries of the chemical species and the reference values of their relative energies are provided in Ref.~\cite{wang2023semilocal}. The inset displays the WTMAD-2, as defined by Eq.~\eqref{wtmad2}, for the TEST12 database obtained using the various functionals under comparison. Note that the assessment of ML-B3LYP-g on the ISO-C dataset in Ref.~\cite{wang2023semilocal} erroneously included the isomerization energy between a species and itself, resulting in an inaccurate MAD. This issue has been corrected here.}
  \label{fig4}
\end{figure*}

It is important to emphasize that, because the training and validation sets exclusively include the absolute energies of atoms and molecules, the prediction and generalization capabilities of the ML-B3LYP-p functional do not rely on error cancellation among different chemical species.

\subsection{Testing performance} \label{subsec:test}

We proceed to evaluate the performance of our optimal NN model by applying it to correct the B3LYP functional and testing it on two databases: the TEST12 database, previously examined in Ref.~\cite{wang2023semilocal}, and the more comprehensive GMTKN55 database.

Figure~\ref{fig4} shows the performance of the ML-B3LYP-p functional on the TEST12 database in comparison to the B3LYP and ML-B3LYP-g methods; see Tables~S1--S5 in the Supporting Information (SI) for details. 
Remarkably, the ML-B3LYP-p functional, trained on pointwise reference data exclusively from absolute energies, achieves the highest accuracy for 9 out of the 12 subsets, which represent different types of relative energies. This demonstrates the feasibility and advantages of constructing an ML-based DFA that does not rely on error cancellation.

The WTMAD-2 for the entire TEST12 database decreases considerably from 43.5 kcal/mol for the original B3LYP functional to 27.8 kcal/mol for the ML-B3LYP-p functional. This value is even lower than the WTMAD-2 for the previously developed ML-B3LYP-g functional, which was trained on both absolute and relative energies \cite{wang2023semilocal}, as shown in the inset of Fig.~\ref{fig4}. The relatively high WTMAD-2 values observed are attributed to the large ratio of $\overline{\left| \Delta E \right|} / \overline{\left| \Delta E \right|}_j$ for certain subsets, such as HTBH38 and ISO20. The lowest WTMAD-2 indicates that ML-B3LYP-p uniformly outperforms the original B3LYP functional, and the ML model developed in this work is superior to its predecessor established in Ref.~\cite{wang2023semilocal}.

For most of the subsets displayed in Fig.~\ref{fig4}, the MADs for the relative energies predicted by the ML-B3LYP-p functional are below 4.0 kcal/mol, consistent with the MAD of 2.5 kcal/mol for the absolute energies of molecules in the G2/97 set predicted by ML-B3LYP-p. However, there are four notable exceptions: P6-AE, AlkAtom19-AE, ISO-C, and BDE42. Specifically, both the AlkAtom19-AE and BDE42 subsets include linear alkane molecules, whose energies are greatly affected by NCIs between pairs of alkane groups \cite{wodrich2006systematic, johnson2012density}. Similarly, the P6-AE and ISO-C subsets contain many cage-like molecules, whose energies are influenced by steric effects, another form of NCIs \cite{johnson2010revealing}. 

The emergence of these exceptions can be attributed to two main factors. First, the piecewise ML correction developed in this work does not directly account for NCIs, as our NN model relies solely on semilocal density descriptors. The training and validation datasets primarily consist of small molecules (mostly hydrides) where NCIs are negligible \cite{julia2011nciplot, tu2023neural}. Second, since our ML correction does not depend on error cancellation, errors related to NCIs may become more pronounced when other errors, such as those associated with covalent bonding, are corrected. 

\begin{figure*}[t]
  \centering
  \includegraphics[width=\textwidth]{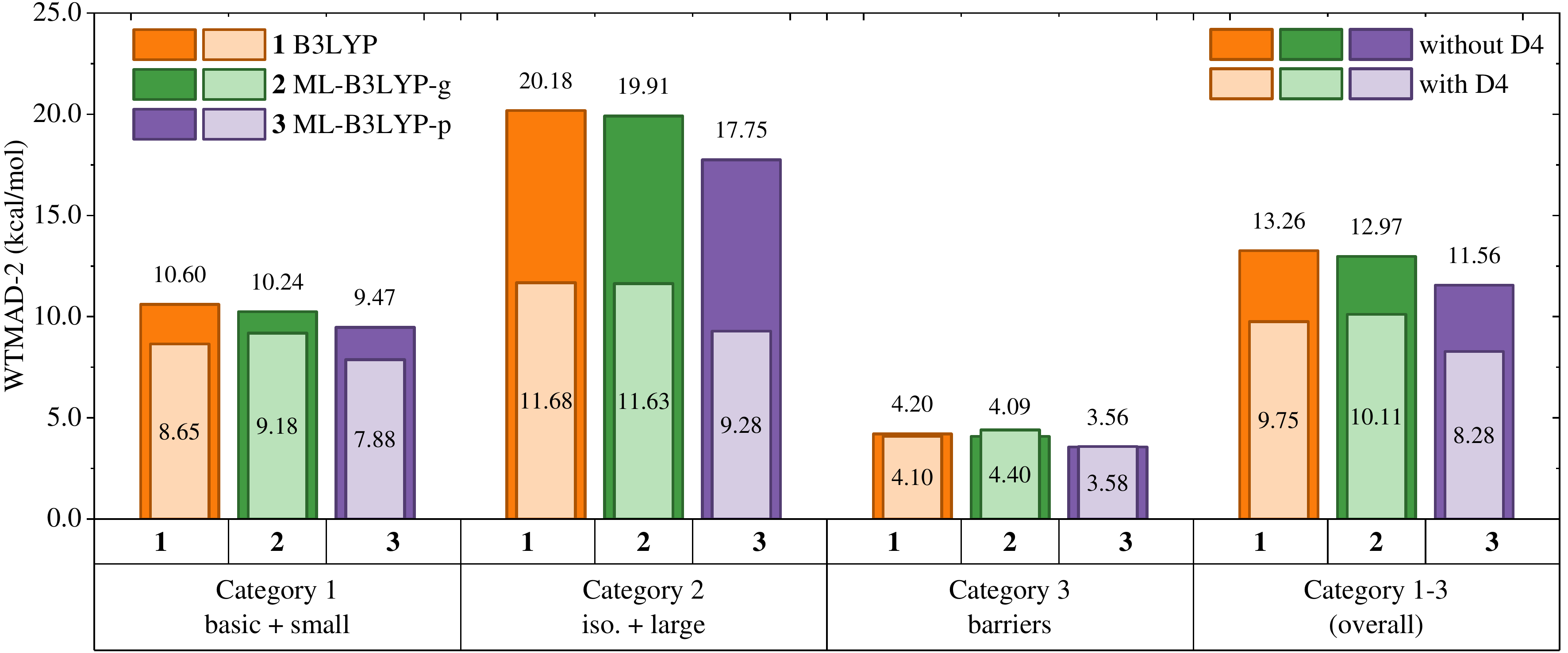} 
  \caption{Performance of the ML-B3LYP-p functional on Categories 1-3 of the GMTKN55 database compared to the B3LYP and ML-B3LYP-g methods. The WTMAD-2 values are presented with and without the D4 correction for NCIs \cite{caldeweyher2017extension, caldeweyher2019generally}; see the main text for further details. The def2-TZVPD basis set is used for elements from from H to Kr, while for subsets containing heavier elements, such as HEAVYSB11, HEAVY28, and HAL59, the def2-ECP basis set \cite{weigend2005a} is employed to treat the core electrons of those heavy atoms.}
  \label{fig5}
\end{figure*}

To assess the generalization ability of ML-B3LYP-p, it is further tested on the comprehensive GMTKN55 database \cite{goerigk2017look}, which is widely used for benchmarking DFAs. Figure~\ref{fig5} compares the WTMAD-2 values for ML-B3LYP-p with those for the B3LYP and ML-B3LYP-g functionals across Categories 1-3 of the GMTKN55 database.

It is somewhat surprising to find that, while ML-B3LYP-g yields slightly lower WTMAD-2 values than the original B3LYP functional without an empirical dispersion correction for NCIs, this minor advantage disappears when the D4 approach \cite{caldeweyher2017extension, caldeweyher2019generally} is applied to account for NCIs. In contrast, the ML-B3LYP-p developed in this work exhibits a consistent and conspicuous reduction in WTMAD-2 values across Categories 1-3, both with and without the D4 dispersion correction. This highlights the excellent generalization capability of our new ML model, which results in uniformly enhanced predictive power across various types of thermochemical and kinetic energies of chemical species.

To further explore the generalization abilities of the ML models, we examine closely the MADs of the 34 constituent subsets in Categories 1-3; see Table~S6 in the SI. For the convenience of discussion, we define a nontrivial improvement in predictive power for a subset as a reduction in MAD of more than 0.5 kcal/mol due to the ML correction. Specifically, ML-B3LYP-g shows nontrivial improvement over the original B3LYP for only 5 subsets, while for the others, it yields similar or even higher MAD values. Moreover, with the D4 approach, the improvements observed in these few subsets are negated. In contrast, ML-B3LYP-p consistently yields lower MADs than the original B3LYP for the majority of subsets in Categories 1-3, with 15 subsets demonstrating nontrivial improvements. These improvements persist with the application of the D4 approach.

The unsatisfactory generalization ability of the ML-B3LYP-g functional is attributed to the limited relative energy data used to represent the chemical space and train the NN model. In this work, the abundance of pointwise data samples that adequately represent the space of electron density significantly enhances the generalization ability of the new NN model incorporated in the ML-B3LYP-p functional. However, there are still a number of subsets for which ML-B3LYP-p does not yield reduced MADs. This is likely due to the limited types of covalent bonds sampled in the training set, as well as the absence of nonlocal density descriptors needed to effectively capture NCIs.

\begin{figure}[t]
  \centering
  \includegraphics[width=\columnwidth]{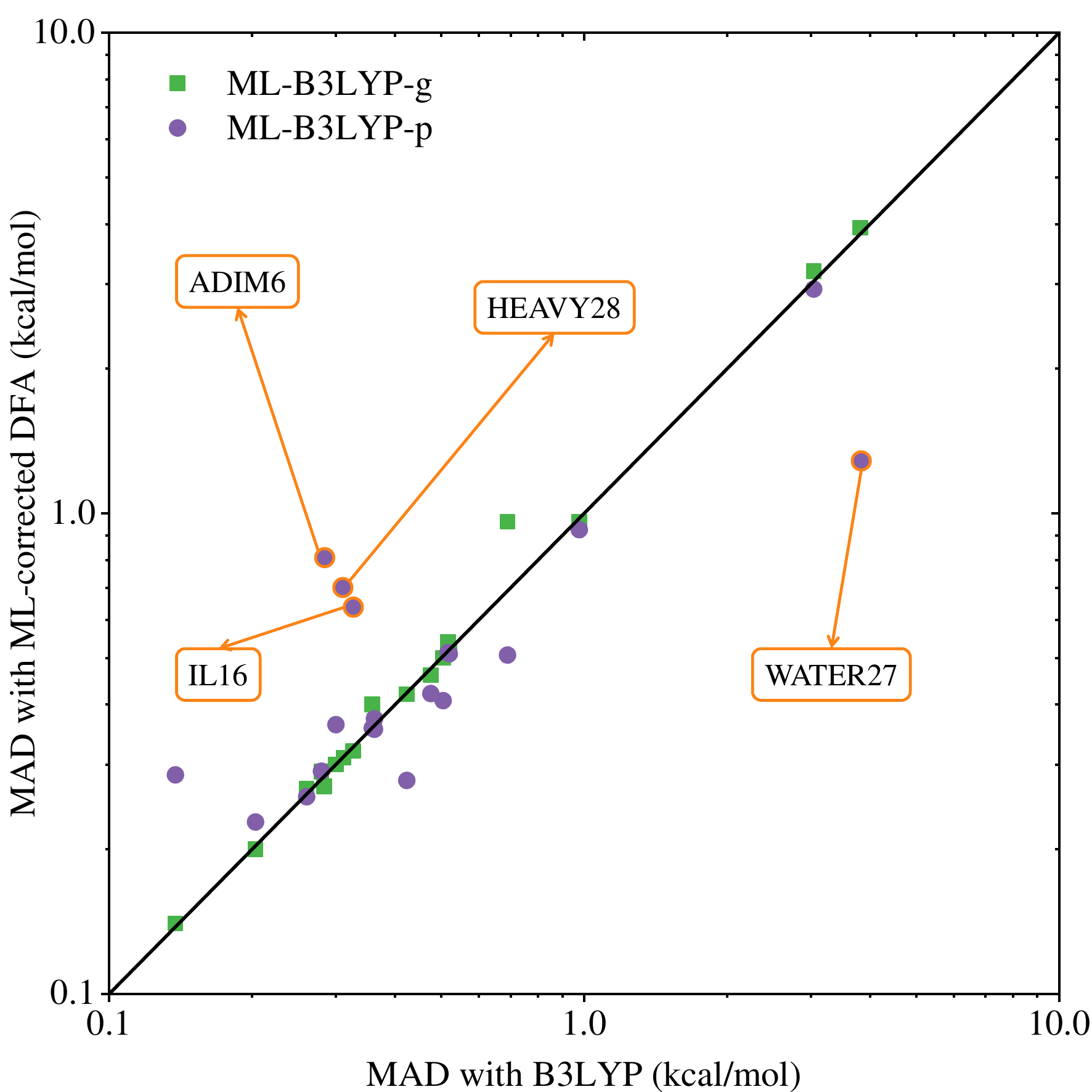} 
  \caption{Scatter plot showing the MADs of the 21 subsets in Categories 4 and 5, which focus on NCIs. The horizontal and vertical axes represent the MAD values associated with the original and ML-corrected B3LYP functionals, respectively. A symbol located above (below) the diagonal line indicates that the ML correction worsens (enhances) the accuracy for the corresponding subset. The D4 correction approach is applied to all DFAs under comparison.}
  \label{fig6}
\end{figure}

Since the ML models in ML-B3LYP-p and ML-B3LYP-g are developed based on a semilocal mapping, they are expected to have a limited impact on the accuracy for Categories 4 and 5, which focus on intermolecular and intramolecular NCIs, respectively. Indeed, as shown in Fig.~\ref{fig6}, the MADs associated with ML-B3LYP-g for the 21 subsets in Categories 4 and 5 are nearly identical to those of the original B3LYP. While ML-B3LYP-p yields substantially reduced MADs for several subsets (such as WATER27), it also results in slightly diminished accuracy for certain subsets (such as ADIM6, HEAVY28, and IL16). Overall, the WTMAD-2 increases from 2.13 kcal/mol to 3.03 kcal/mol for B3LYP and ML-B3LYP-p  for Category 4, while it slightly decreases from 0.52 kcal/mol to 0.51 kcal/mol for Category 5.

\begin{figure}
    \centering
    \includegraphics[width=\linewidth]{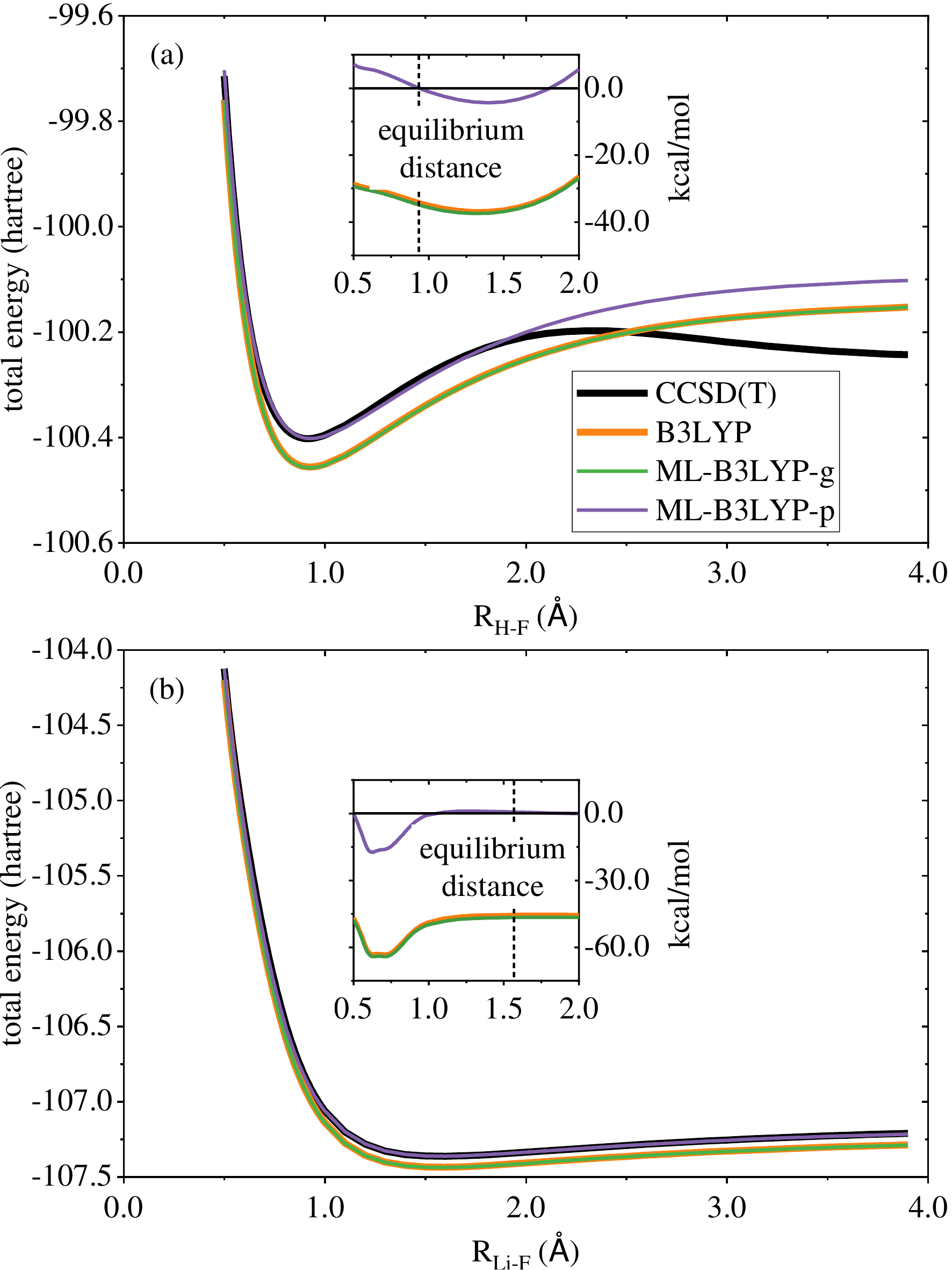}
    \caption{Dissociation energy curves of (a) HF and (b) LiF molecules calculated using the ML-B3LYP-p functional, compared to those obtained from the original B3LYP and ML-B3LYP-g functionals. The reference data calculated using the CCSD(T) method are also displayed. The insets show the deviations of the density functional methods from CCSD(T) for bond lengths ranging from 0.5 to 2.0 \si{\angstrom}. The dashed vertical lines indicate the equilibrium bond lengths: 0.93 \si{\angstrom} for HF and 1.57 \si{\angstrom} for LiF.}
    \label{fig7}
\end{figure}

The NN model incorporated in ML-B3LYP-p is trained on molecules at equilibrium structures. To evaluate its performance on chemical species in nonequilibrium configurations, we examine the dissociation energy curves of two diatomic molecules, HF and LiF, within a region around their equilibrium bond lengths, as shown in Fig.~\ref{fig7}. In both cases, the energy curves obtained from ML-B3LYP-p closely overlap with those from CCSD(T) when the bond lengths are shorter than 2.0 \si{\angstrom}. At longer bond lengths, the curves from ML-B3LYP-p run parallel to those from B3LYP, maintaining a nearly constant difference between them.

The comparisons shown in Fig.~\ref{fig7} suggest that the effectiveness of the ML correction persists when a chemical bond is moderately compressed or stretched. This explains the enhanced accuracy in predicting transition state energies and reaction barriers, as indicated by the reduced MADs for the HTBH38 and NHTBH38 subsets in the TEST12 database and Category 3 in the GMTKN55 database. However, when bond lengths are stretched to very long distances, approaching the dissociation limit, the results from ML-B3LYP-p deviate significantly from those of CCSD(T). These deviations arise from severe delocalization and static correlation errors \cite{Cohen2012challenges, cohen2008insights}, which are not represented in our current training set.

\subsection{Computational Efficiency} \label{subsec:efficiency}

\begin{figure}[t]
    \centering
    \includegraphics[width=\linewidth]{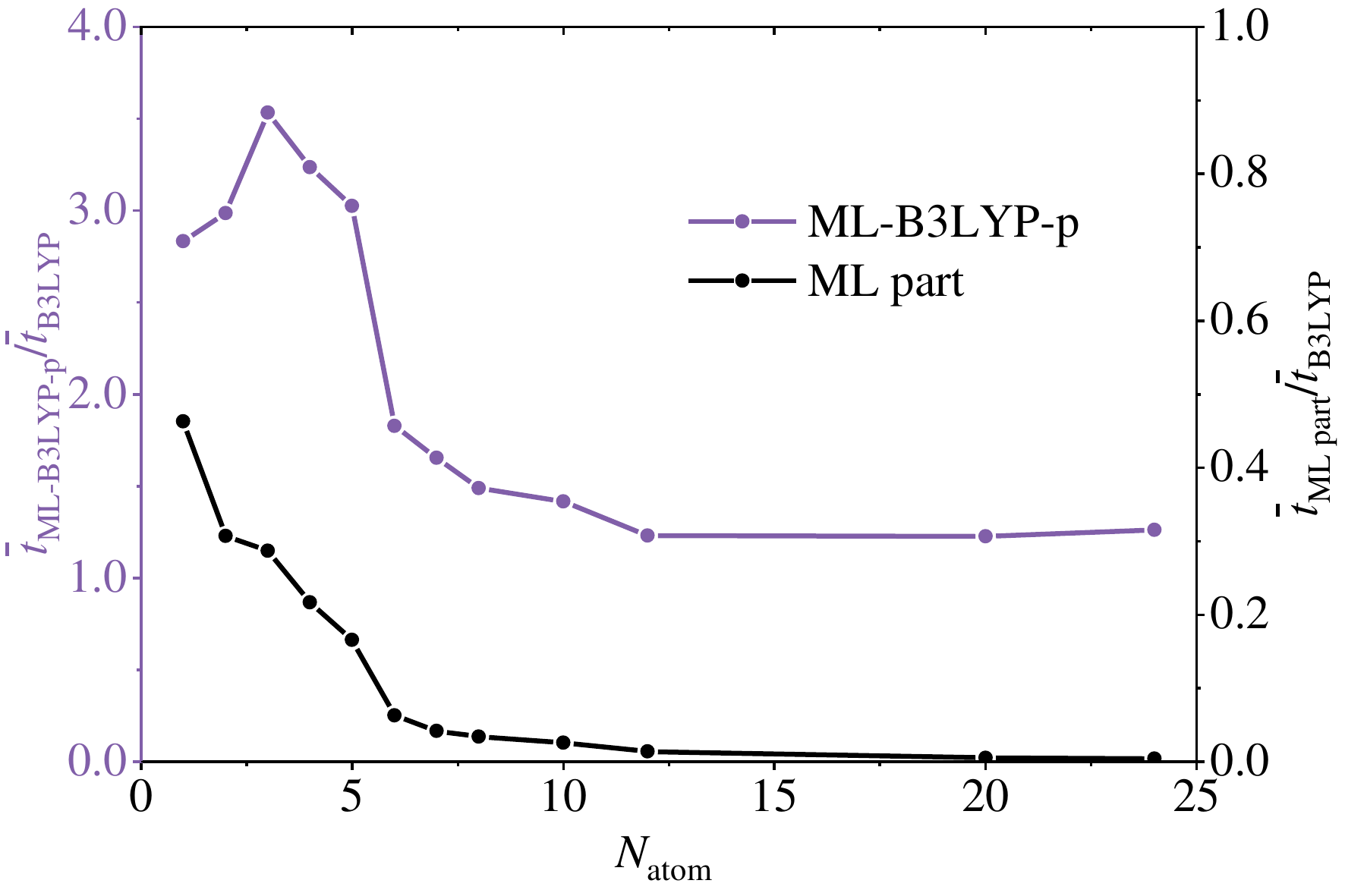}
    \caption{The ratio of the average computational time required for ML-B3LYP-p to that for B3LYP in single-point calculations for chemical species of varying sizes in the TEST12 database; see the main text for further details. The horizontal axis represents the sizes of species, quantified by the number of non-hydrogen atoms in each species, $N_{\rm atom}$.}
    \label{fig8}
\end{figure}

To establish an ideal substitute for the B3LYP functional, the computational cost of the ML-corrected B3LYP method must remain comparable to that of B3LYP. To assess the computational time required for ML-B3LYP-p in relation to B3LYP, we conduct a statistical analysis of the relative computational time, characterized by the ratio $\bar{t}_{\rm ML-B3LYP-p}/\bar{t}_{\rm B3LYP}$ for chemical species in the TEST12 database, and plot the ratio as a function of $N_{\rm atom}$ in Fig.~\ref{fig8}. Here, $\bar{t}_{\rm ML-B3LYP-p}$ and $\bar{t}_{\rm B3LYP}$ represent the average computational times needed to calculate the single-point energies of chemical species containing $N_{\rm atom}$ non-hydrogen atoms using the ML-B3LYP-p and B3LYP methods, respectively.

As depicted in Fig.~\ref{fig8}, the ratio $\bar{t}_{\rm ML-B3LYP-p}/\bar{t}_{\rm B3LYP}$ decreases as $N_{\rm atom}$ increases, approaching a constant value of 1.3 in the large-$N_{\rm atom}$ limit. This suggests that the computational time required for ML-B3LYP-p remains comparable to that of the original B3LYP. Moreover, Fig.~\ref{fig8} exhibits the ratio $\bar{t}_{\rm ML-part}/\bar{t}_{\rm B3LYP}$ as a function of $N_{\rm atom}$, where $\bar{t}_{\rm ML-part}$ represents the average time consumed by PyTorch for calculating the ML correction to the XC energy and Kohn-Sham Hamiltonian. Apparently, this ratio rapidly decays to zero as $N_{\rm atom}$ increases, indicating that the time required for the ML correction becomes negligible compared to that of the SCF process for large chemical species.

In addition to accessing the NN model for calculating $\Delta E^{\rm ML}_{\rm XC}$ and $\Delta v^{\rm ML}_{\rm XC}$, the introduction of the ML correction also requires the calculation of the kinetic energy density $\tau$. Moreover, the incorporation of the ML correction may affect the convergence process of SCF cycles. These additional procedures result in a 30\% increase in computational time for large $N_{\rm atom}$.

\section{Conclusion and Perspective} \label{sec:conclusion}

In this work, we successfully constructed a novel NN model based on a semilocal mapping, trained on an extensive collection of real-space pointwise data derived from highly accurate absolute energies. This thus eliminates the need for error cancellation to enhance the predictive capabilities of the target DFA. We introduced a double-cycle training protocol that enables efficient training using gradient-based optimizers. The resulting ML-B3LYP-p model demonstrates an excellent balance between accuracy and computational efficiency, outperforming the original B3LYP functional and the previously developed ML-B3LYP-g functional. These achievements position ML-B3LYP-p as a superior alternative in the realm of DFT methods, making it particularly well-suited for general-purpose applications across a wide range of chemical systems.

Our proposed NN model and training strategy can be readily extended to other types of DFAs. It is expected that applying the ML correction approach to a parent DFA that is more sophisticated than B3LYP will yield an even more advanced ML-corrected functional. This  transferability underscores the versatility of our methodology and its potential to enhance the performance of various DFAs beyond those demonstrated in this work.

Nonetheless, several aspects of our study require further optimization. First, the NN in ML-B3LYP-p currently lacks nonlocal density descriptors, which limits its effectiveness in accurately describing NCIs. Nagai et al. introduced an $R$-function that convolutes the electron density in the vicinity of a point as a descriptor, enhancing the predictive accuracy of ML-based DFAs \cite{nagai2020completing}. However, integrating the $R$-function into the SCF process remains computationally expensive. This accentuates the need for further exploration and improvement in the design and application of nonlocal density descriptors. Second, our training set is restricted to small molecules. While this has led to some advancements, incorporating a broader range of molecular types could yield even better training outcomes. Developing a more efficient method for acquiring accurate pointwise reference data is essential for accommodating larger molecules in the training process. Finally, the current NN model is a relatively simple multilayer perceptron. Exploring more advanced NN architectures may further enhance the model's capability to represent the complex mapping from electron density to XC energy.

In addition to ML-based corrections, the rational design of physics-inspired methods to address well-known systematic errors in DFAs \cite{Cohen2012challenges, cohen2008insights, perdew1985density}, such as delocalization error, static correlation error, and self-interaction error, is also crucial. By combining these two approaches, we may ultimately attain an XC functional that universally achieves chemical accuracy.

\section*{Acknowledgement} \label{sec:acknow}

Support from the National Natural Science Foundation of China (Grant Nos. 22393912, 22321003, 22425301), the AI for Science Foundation of Fudan University (Grant No. FudanX24AI023), the Strategic Priority Research Program of the Chinese Academy of Sciences (Grant No. XDB0450101), and the Hong Kong Quantum AI Lab, AIR@InnoHK of the Hong Kong Government is gratefully acknowledged.

\section*{Data Availability} \label{sec:data_ava}

The authors confirm that the data and code supporting the findings of this study are available within the supplementary materials:

\begin{itemize}
    \item \textbf{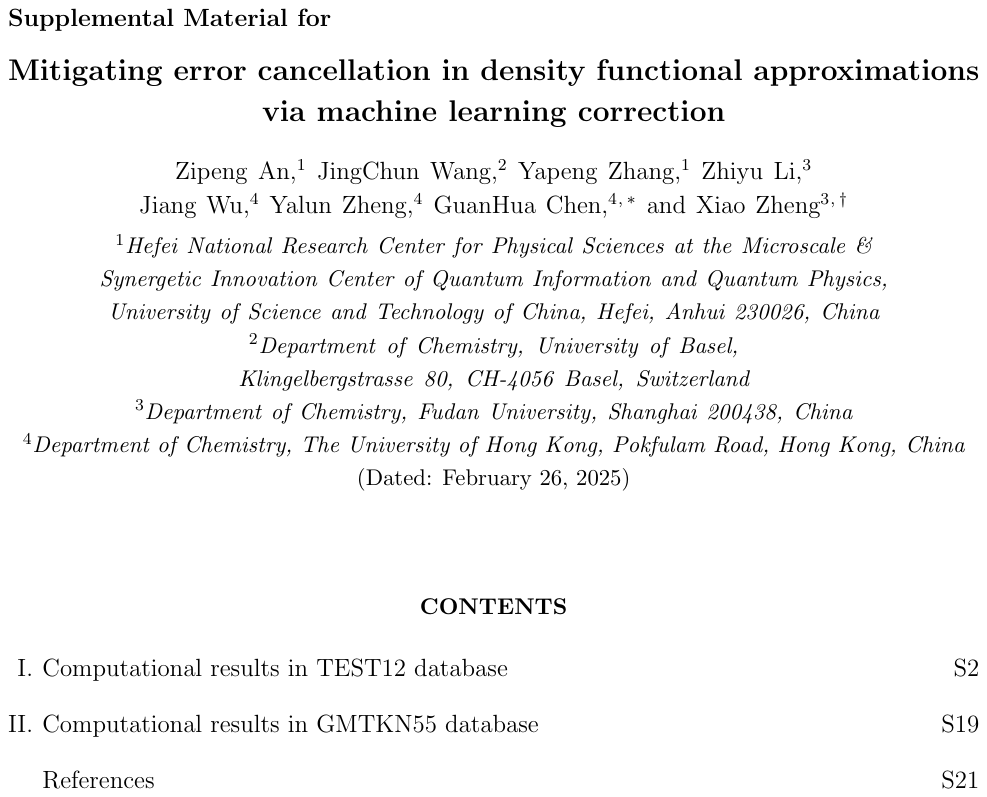}: 
    Computational details on the benchmark tests for the ML-B3LYP-p, ML-B3LYP-g, and B3LYP functionals.
    
    \item \textbf{database.zip}: 
    Raw data and processing scripts for the benchmark tests of ML-B3LYP-p, ML-B3LYP-g, and B3LYP.
    
    \item \textbf{Source Code}: 
    Computer codes that implement the ML framework and analysis tools developed in this work. \\
    GitHub repository: \url{https://github.com/ZipengAn/ML-B3LYP-p}
\end{itemize}


\bibliography{bibrefs}

\end{document}